\documentclass[a4paper]{article}

\usepackage[pages=all, color=black, position={current page.south}, placement=bottom, scale=1, opacity=1, vshift=5mm]{background}
\SetBgContents{
	
}      % copyright

\usepackage[margin=1in]{geometry} % full-width

\usepackage{amsmath}
\usepackage{amsthm}
\usepackage{amssymb}
\usepackage{hyperref}
\hypersetup{
	unicode,
	pdfauthor={Shayan Roofeh, Vahid Karimipour},
	pdftitle={Capacities of Generalized Landau-Streater Channel},
	pdfsubject={Capacities of Generalized Landau-Streater Channel},
	pdfkeywords={article, template, simple},
	pdfproducer={LaTeX},
	pdfcreator={pdflatex}
}

\usepackage[sort&compress,numbers,square]{natbib}
\usepackage{bm}
\theoremstyle{plain}

\theoremstyle{definition}

\usepackage{graphicx}
\graphicspath{{fig/}}
\usepackage{subcaption}
\usepackage{algorithm, algpseudocode} % use algorithm and algorithmicx for typesetting algorithms
\usepackage{mathrsfs} % for \mathscr command
\usepackage{lipsum}
\usepackage{physics}
\def\be{\begin{equation}}
	\def\ee{\end{equation}}
\def\ba{\begin{eqnarray}}
	\def\ea{\end{eqnarray}}
\def\lo{\longrightarrow}
\def\h{\hskip 1cm }

\def\la{\langle}
\def\ra{\rangle}
\def\a{\alpha}

\def\ni{\noindent}
\def\bex{\begin{dinglist}{110}\dsquare}
	\def\eee{\end{dinglist}}
\def\bet{\begin{dinglist}{110}\bsquare}
	\def\bfr{\begin{mdframed}[backgroundcolor=blue!20]\vspace{0.5cm}}
		\def\efr{\vspace{0.5cm}\end{mdframed}}
	
	\title{ Capacities of a two-parameter family of noisy Werner-Holevo channels}
	\author{ Shayan Roofeh and Vahid Karimipour}
	
	\date{
		Deptartment of Physics, Sharif University of Technology, Tehran, Iran\\%
	}
	
\begin{document}
		\maketitle
		
		\begin{abstract}
			In $d=2j+1$ dimensions, the Landau-Streater quantum channel is defined on the basis of spin $j$ representation of the $su(2)$ algebra. Only for $j=1$, this channel is equivalent to the Werner-Holevo channel and enjoys covariance properties with respect to the group $SU(3)$.  We extend this class of channels  to higher dimensions in a way which is based on the Lie algebra $so(d)$ and $su(d)$. As a result it retains its equivalence to the Werner-Holevo channel in arbitrary dimensions. The resulting channel is covariant with respect to the unitary group $SU(d)$. We then modify this channel in a way which can act as a noisy channel on qudits. The resulting modified channel now interpolates between the identity channel and the Werner-Holevo channel and its covariance is reduced to the subgroup of orthogonal matrices $SO(d)$.  We then investigate some of the propeties of the resulting two-parameter family of channels, including their spectrum, their regions of lack of indivisibility, their Holevo quantity,  entanglement-assisted capacity and the closed form of their complement channel and a possible lower bound for their quantum capacity.  
			
%			\noindent\textbf{Keywords:} 
		\end{abstract}
	\section{Introduction} 
	
	The Werner-Holevo channel, defined as
	\be
	\Phi_{WH}(\rho)=\frac{1}{d-1}\big(Tr(\rho)I-\rho^T\big)
	\ee
		 was first introduced in \cite{WH} and since then has 
		 attracted a lot of attention as examples of quantum channels for which the output purity is not additive. When acting on three-dimensional states or qutrits, these channels are equivalent to the Landau-Streater channel, defined as 
		 \be
		 \Lambda_{LS}(\rho)=\frac{1}{2}\big(J_x\rho J_x+J_y\rho J_y+J_z\rho J_z\big)
		 \ee
	where $J_x, \ J_y$ and $J_z$ are the spin-1 representation of the angular momentum operators. Many aspects of these channels have been extensively studied in recent years. An important property of this channel in three dimensions (Werner-Holevo or Landau-Streater) is that it is an extreme point in the space of quantum channels and cannot be represented as the convex combination of two other channels. Quite recently, however we could define a one-parameter extension of these qutrit channels in the form 
	\be
	\Lambda_x(\rho)=(1-x)\rho + \frac{x}{2}(Tr(\rho)I-\rho^T)
	\ee
	 and call them noisy Werner-Holevo channel \cite{roofeh_noisy_2023}, where for a large portion of the noise parameter $x$ the channel was shown to be indeed a convex combination of random unitary channels. In fact we could show \cite{roofeh_noisy_2023} that the noisy channel has a nice and simple physical expression, namely that the qutrit state $\rho$, when passing through the channel receives random kicks in the form of random rotations around random axes and turns into $\Lambda_x(\rho)$. In \cite{roofeh_noisy_2023} various bounds for different capacities of this family of channels was derived and more importantly it was shown that when the level of noise $x$ is above $\frac{4}{7}$, the channel is anti-degradable and hence its quantum capacity vanishes. In \cite{lo_degradability_2024} the Landau-Streater channel for arbitrary spin was studied in terms of epsilon-degradability. Finally quite recently it was shown in \cite{karimipour_noisy_2024} that the equivalence between the two channels, namely the Werner-Holevo and the Landau-Streater channel can be maintained if we formulate the latter channel for the group$O(d)$. After studying capacities for a one-parameter noisy extension of these channels, it was shown in \cite{karimipour_noisy_2024} that when the dimension is even, these channels are a convex combination of unitary operations, a property which is absent in odd dimensions. \\

	 \ni In this paper we extend the work of \cite{karimipour_noisy_2024} to a two-parameter family of channels and explain the extension to the unitary groups $SU(d)$ and $U(d)$. We then calculate upper and lower bounds for different capacities of the channel, namely the classical capacity, the entanglement-assisted capacity and the quantum capacity. These bounds depend on the dimension and the two parameters of the channel. In particular, for the quantum capacity we determine  the region of parameters where the quantum capacity vanishes exactly. This region is of course not the entire region of zero capacity but is a rather significant part of the parameter space.  \\
	 
	\ni The structure of this paper is as follows: In section (\ref{Preliminaries}), we introduce preliminary materials which are needed for the rest of the paper. Furthermore, we introduce the two parameter family of channels that we study in this paper, in section (\ref{classicalcapacity}), we calculate the Holevo quantity of the channel and set a lower and an upper bound for the classical capacity. In section (\ref{ent}), the entanglement-assisted classical capacity is determined. Finally in section (\ref{QuantumCapacity}), we determine the upper bound, the lower bound and the zero-capacity region of the quantum capacity. We end with a conclusion.

	\iffalse

	In the space of quantum channels, two specific quantum channels  have widely attracted the attention of researchers, due to their specific properties. These are the Landau-Streater (LS) channel \cite{LanS} which we denote by ${\cal L}_j$ and the Werner-Holevo channel \cite{WH}, denoted by $\phi_{_{\eta,d}}$ \cite{WH}. Various properties of the Landau-Streater channel have been studied for example in \cite{Audenaert_2008, filippov_LS, pakhomchik_realization_2020} and those of the Werner-Holevo channel in \cite{ Girard_2022, datta2004additivity, Cope_2017, Cope_2018, Chitambar_2023, Wolf_2005}.	 Below we will remind the reader of their definition.\\
	
	\fi
	
	\section{Preliminaries}
	\label{Preliminaries}
	The space of qudits is denoted by $C_d$ with basis states $\{|m\ra,\ \ m=1\cdots d\}$, a set of generators of the Lie algebra of $su(d)$ is chosen to be Gell-mann matrices, defined as
	\ba
	J_{mn}&=&-i(|m\ra\la n|-|n\ra\la m|)\h  1\leq m<n\leq d\cr
	K_{mn}&=&|m\ra\la n|+|n\ra\la m|,\h\ \ \ \  1\leq m<n\leq d \ea
	and
	\be
	K_m=\sqrt{\frac{2}{m(m+1)}}\big[\sum_{i=1}^{m}|i\ra\la i|-m|m+1\ra\la m+1|\big]\h 1\leq m\leq d-1
	\ee
Let us denote these three sets respectively by ${\cal A}^-, {\cal A}^+$ and ${\cal A}^0$. The union of these sets comprise  the $d^2-1$ traceless-Hermitian matrices which are orthogonal to each other in the sense that $Tr(A_\a A_\beta)=2\delta_{\alpha, \beta}$. Now let $\rho=\sum_{m,n}\rho_{m,n}|m\ra\la n|$, be a general $d-$ dimensional density matrix. Then it is a matter of straightforward calculation to define three different kind of quantum channels (i.e. Completely Positive Trace Preserving) maps by adopting each of the above three sets as Kraus operators of that channel. One finds

	\be
\Phi^-(\rho):=\frac{1}{d-1} \sum_{m<n}J_{mn}\rho J_{mn}=\frac{1}{d-1}\big(\tr(\rho)I - \rho^T\big)
\label{Channel1}
\ee
	\be
\Phi^+(\rho):=\frac{1}{d-1} \sum_{m<n}K_{mn}\rho K_{mn}=\frac{1}{d-1}\big(\rho^T+\tr(\rho)I - 2\rho_{_D}\big),
\label{Channel2}
\ee
and
	\be
	\label{DecoheringChannel}
\Phi^0(\rho)=\frac{d}{2(d-1)}\sum_{m=1}^{d-1} K_m\rho K_m=\frac{d}{d-1}\big(\rho_{_D}-\frac{1}{d}\rho\big).
\ee

\ni where $\rho_D=\sum_n\rho_{nn}|n\ra\la n|$ is the completely decohered state of $\rho$. The coefficients have been inserted to guarantee preservation of trace. In view of the simple form of the Kraus operators $J_{mn}$ and $K_{mn}$, verification of the first two equations is straightforward. We provide a proof for the third equation. \\

\ni The proof is by induction on $d$. For clarity and only for this proof, we denote a Hermitian matrix of dimension $d$ by $X^{(d)}$. Equation (\ref{DecoheringChannel}), or equivalently 
	\be
 \sum_{m}K_{m}X^{(d)} K_{m}=2\big(X^{(d)}_D - \frac{1}{d}X^{(d)}\big),
\label{Channel2}
\ee
 clearly holds for $d=2$. Assume that it holds for $d$ and note that $$K_d=\sqrt{\frac{2}{d(d+1)}}\begin{pmatrix}I_d&{\bf 0}\\ {\bf 0}^T& -d\end{pmatrix}.$$
One then finds that, in view of the Block structure of all the operators $K_m$,  the set of Kraus operators in $A^0$ act as follows on a general Hermitian matrix like $$ X^{(d+1)}=\begin{pmatrix}X^{(d)},&{\bf r}\\ {\bf r}^\dagger& s\end{pmatrix}:$$ 
\ba\nonumber
\sum_{m=1}^{d} K_m X^{(d+1)} K_m&=&\begin{pmatrix}2 X^{(d)}_{_D}-\frac{2}{d}X^{(d)},&{\bf 0}\\ {\bf 0}& 0\end{pmatrix}+K_d X^{(d+1)}K_d\cr &=&\begin{pmatrix}2 X^{(d)}_{_D}-\frac{2}{d}X^{(d)},&{\bf 0}\\ {\bf 0}& 0\end{pmatrix}+\frac{2}{d(d+1)}\begin{pmatrix}I_d,&{\bf 0}\\ {\bf 0}& -d\end{pmatrix}\begin{pmatrix} X^{(d)},&{\bf r}\\ {\bf r}^\dagger& s\end{pmatrix}\begin{pmatrix}I_d&{\bf 0}\\ {\bf 0}^T& -d\end{pmatrix}
\ea
which after simplification becomes
\be
\sum_{m=1}^{d} K_m X^{(d+1)} K_m=\begin{pmatrix}2 X^{(d)}_{_D}-\frac{2}{d+1} X^{(d)},&-\frac{2}{d+1} {\bf r}\\ -\frac{2}{d+1} {\bf r^\dagger} &(2-\frac{2}{d+1})s\end{pmatrix}=2 \big(X^{(d+1)}_{_{D}}-\frac{1}{d+1} X^{(d+1)}\big).
\ee

This completes the proof.\\

\ni Instead of the set $A^+$, one can also adopt a different set of Kraus operators consisting of the following Hermitian operators 
\be
{\cal A}^{(+)}:=\{K_{mn}=|m\ra\la n|+|n\ra\la m|,\ \ \ \ \  1\leq m,n\leq d\}
\ee
The operators in this set, i.e. the ones with $m=n$ are no longer trace less and indeed the union of ${\cal A}^-\cup {\cal A}^{(+)}$ generate the Lie-Algebra of the unitary group $U(d)$ with a total of $d^2$ elements. Interestingly the set ${\cal A}^{(+)}$ generates a Werner Holevo channel without the decohered part $\rho_D$:

	\be
\Phi^{(+)}(\rho)=\frac{d}{(d+1)}\sum_{K_{m,n}\in {\cal A}^{(+)}} K_{m,n}\rho K_{m,n}=\frac{1}{d+1}\big(Tr(\rho)I +\rho^T\big).
\ee
	One can make a convex combination of the channels in (\ref{Channel1}), (\ref{Channel2}) and (\ref{DecoheringChannel}) in different ways in order to make a channel whose output is a combination of $\rho, \ \rho^T, \ \rho_{{_D}} $ and the completely mixed state $\frac{I}{d}$ with arbitrary coefficients. Since $\rho_{_D}$ and $\rho^T$ appear in these three channels with alternating signs, one can also make combinations in which either $\rho^T$ or $\rho_{_D}$ are absent. \\
	
	\ni In this work we analyze the following channel 
	
	\be
	\label{Two-Parameter-Channel}
	\Phi_{_{x,y}}(\rho)=(1-x-y)\rho + x \Phi^{-}(\rho)+ y\Phi^{(+)}(\rho), 
	\ee
	where $x$ and $y$ are two positive real parameters subject to the condition $x+y\leq 1$.
which acts as follows
\be
	\Phi_{_{x,y}}(\rho)=(1-x-y)\rho+(y_d+x_d)Tr(\rho)I+(y_d-x_d)\rho^T
\ee
in which we have defined two new parameters:
\be
x_d:=\frac{x}{d-1},\h y_d:=\frac{y}{d+1}.
\ee	
The reason for this definition, as we will see in the sequel, is that these two new parameters appear over and over again in various formulas for capacities.	
\noindent	It is noteworthy to explain the relation of the channel (\ref{Two-Parameter-Channel}) and the well-known Werner-Holevo \cite{WH} channel which is usually defined and parameterized in the form \cite{cope_adaptive_2017}:

\be
\phi_{{_{\eta,d}}}(\rho):= \frac{1}{d^2-1}\left[(d-\eta)\tr\rho\  \mathcal{I}+(d\eta-1)\rho^T)\right]
\ee
If we set $x=\frac{1-\eta}{2}$ and $y=\frac{1+\eta}{2}$, the Werner-Holevo channel is retrieved. Thus the Werner-Holevo channel is a special case of the channel $\Phi_{_{x,y}}$, when we $x+y=1$ and when $x+y<1$, we can say that we have a noisy Werner-Holevo channel, meaning that the identity channel is affected by a noise which is the Werner-Holevo channel. The pure Werner-Holevo channel is covariant under the group $U(d)$, since 
\be
\phi_{_{\eta,d}}(U\rho U^\dagger)=U^*\phi_{_{\eta,d}}(\rho)U^T,\h \forall\ U\in U(d),
\ee
a property which facilitates many calculations relating to the capacities of quantum channels. However the channel $\Phi_{x,y}$ has a much lower symmetry, namely 
\be
\phi_{x,y}(U\rho U^\dagger)=U^*\phi_{_{\eta,d}}(\rho)U^T,\h \forall\ U\in O(d),
\ee	
where $O(d)$ is the orthogonal group in $d-$ dimensions. As we will see these symmetries, will play essential role in calculating the Holevo quantity and the entanglement-assisted capacity. It is specially important to note that the symmetry depends on whether we are on the line $x+y=1$ or below it, figure (\ref{fig:Phases}). 
 We can now determine the Holevo information of the channel $\Phi_{_{x,y}}$. This is done in the next section. 

\section{Classical Capacity}\label{classicalcapacity}
In this section, we first find the Holevo information which is the a lower bound for classical capacity. As expected the covariance properties of the channel plays a significant role in determining the analytical form of this quantity. It turns out that the Holevo information has a different analytical form on the line $x+y=1$ and $x+y<1$. Therefore we consider these two cases separately.  
\subsection{The Holevo information on the line  $x+y=1$.}
We start from the well-known results \cite{holevo_remarks} which states that for an irreducible quantum channel in $d-$ dimension, 
\be
C^1(\Phi_{_{x,y}})=\log d - min_{\psi} S(\Phi_{_{x,y}}(|\psi\ra\la \psi|))
\ee
where $S$ is the von-Neumman entropy $S(\rho)=-Tr(\rho\log_2\rho)$. Due to the concavity of $S$, the minimum output entropy is a pure state \cite{wilde_book_2017}. Therefore our task is to find the input pure state which maximizes the output entropy. To this end, we use the covariance property 
$$S(\Phi(|\psi\ra)\la \psi|)=S(\Phi(U|\psi_0\ra)\la \psi_0|U)=S(U^*\Phi(|\psi_0\ra)\la \psi_0|U^T)=S(\Phi(|\psi_0\ra)\la \psi_0|)$$
to restrict the search to a single reference state $|\psi_0\ra$ which we take to be $|\psi_0\ra=|1\ra=\begin{pmatrix}1& 0 & 0 &\cdots 0\end{pmatrix}^T$.  One then finds 

\be
\Phi_{_{x,y}}(|\psi_0\ra\la \psi_0|)=(y_d+x_d)\ \mathcal{I} +  (y_d-x_d)\ |1\ra\la 1|.
\ee
The eigenvalues of this output state are 
\be
\{2y_d\ (g=1),\ \ \  (y_d+x_d)\ (g=d-1)  \}
\ee
where multiplicty of each eigenvalue is indicated by $g$. 
The von Neumman entropy of this state is now calculated to be 
\be
C^1(\Phi_{_{x,y}})=\log d +2y_d\log y_d+(d-1)\big(y_d+x_d\big)\log \big(y_d+x_d\big)
\ee
Parameterizing $x$ and $y$ as 
$
y=:\frac{1+\Delta}{2}, $ and $ x=:\frac{1-\Delta}{2}
$,
we find
\be
C^1(\Phi_{_{x,y}})=\log d +\frac{1+\Delta}{d+1}\log \frac{1+\Delta}{d+1}+\frac{d-\Delta}{d+1}\log\frac{d-\Delta}{d^2-1}.
\ee
Note that without using the $U(d)$ covariance of the channel, this result could not have been derived, however this large covariance is present only when we are on the edge of the triangle $x+y=1$.    Inside the triangle, we have a much smaller covariance which is treated separately in the next subsection.

\subsection{The Holevo information inside the triangle, i.e.  $x+y<1$.}
Let the minimum output entropy state be denoted by \( |\psi\rangle \). Since we no longer have the \( U(d) \) covariance, we are unable to transform this state \( |\psi\rangle \) into a chosen reference state. Instead, we adopt a different approach \cite{karimipour_noisy_2024} and explore how far we can simplify the parameters of the state \( |\psi\rangle \) by leveraging the \( O(d) \) covariance. For ease of reference we repeat the reasoning of \cite{karimipour_noisy_2024} and adopt it to the two-parameteric case. Let \( |\psi\rangle \) be represented as  
\[
|\psi\rangle = \begin{pmatrix}
\psi_1 \\
\psi_2 \\
\psi_3 \\
\vdots \\
\psi_d
\end{pmatrix}
\]
where, except for a global phase, all the parameters \( \psi_i \) are complex numbers and must satisfy normalization. We begin by applying a rotation generated by \( J_{12} \), which transforms 
\[ 
\psi_2 \to -\sin\theta \, \psi_1 + \cos\theta \, \psi_2 
\]
to eliminate the imaginary part of \( \psi_2 \), making it real. We will denote this real value as \( r_2 \). By successively applying the covariance generated by \( J_{13} \), \( J_{14} \), \(\cdots\), \( J_{1d} \), we transform all the other coefficients \(\psi_3, \psi_4, \cdots, \psi_d\) into real numbers, thus expressing \( |\psi\rangle \) in the form
\[
|\psi\rangle = \begin{pmatrix}
\psi_1 \\
r_2 \\
r_3 \\
\vdots \\
r_d
\end{pmatrix}, \quad r_i \in \mathbb{R}.
\]
Next, we use rotations generated by \( J_{23}, J_{24}, \cdots, J_{2d} \) to make all the parameters \( r_i \) except \( r_2 \) equal to zero, transforming the state \( |\psi\rangle \) into the form 
\[
|\psi\rangle = \begin{pmatrix}
\psi_1 \\
r_2 \\
0 \\
\vdots \\
0
\end{pmatrix} = \begin{pmatrix}
\cos\theta e^{i\phi} \\
\sin\theta \\
0 \\
\vdots \\
0
\end{pmatrix}.
\]
The output state is then given by
\[
\begin{split}
\Phi_{_{x,y}}(|\psi\rangle \langle \psi|) &= (1-x-y)\begin{pmatrix}
\cos^2\theta & \cos\theta \sin\theta e^{i\phi} \\
\cos\theta \sin\theta e^{-i\phi} & \sin^2\theta
\end{pmatrix} \bigoplus \mathbf{0}^{d-2} \\
&\quad + \begin{pmatrix}
1 & 0 \\
0 & 1
\end{pmatrix} \bigoplus (y_d + x_d) \mathcal{I}^{d-2} \\
&\quad + (y_d - x_d)\begin{pmatrix}
\cos^2\theta & \cos\theta \sin\theta e^{i\phi} \\
\cos\theta \sin\theta e^{-i\phi} & \sin^2\theta
\end{pmatrix} \bigoplus \mathbf{0}^{d-2}.
\end{split}
\]
This implies that the eigenvalues of the output state are given by
\[
\text{Spectrum of } \Phi_{_{x,y}}(|\psi\rangle \langle \psi|) = \{(y_d + x_d), g = d-2\} \cup \text{Spectrum of } M,
\]
where \( g = d-2 \) denotes the multiplicity of the first eigenvalue, and \( M \) is a two-dimensional matrix.
\[
M=\begin{pmatrix}
(1-x-y)\cos^2\theta + y_d(1+\cos^2\theta) + x_d\sin^2\theta & A\cos\theta \sin\theta \\
A\cos\theta \sin\theta & (1-x-y)\sin^2\theta + y_d(1+\sin^2\theta) + x_d\cos^2\theta
\end{pmatrix}.
\]
Here, \( A \) is defined as
\[
A = (1-x-y) e^{i\phi} + (y_d - x_d) e^{-i\phi}.
\]
To find the state that minimizes the output entropy, it is not necessary to explicitly compute the eigenvalues of this matrix. It is enough to note that the trace of the matrix, which is the sum of its eigenvalues, is given by
\[
\label{sumeig}
\text{tr} \, M = \lambda_1 + \lambda_2 = 1 - (d-2)(x_d + y_d),
\]
and is independent of the input state. The determinant, which is the product of the eigenvalues, is given by
\[
\det M = Q(\theta, x, y) Q\left(\frac{\pi}{2} - \theta, x, y\right) - R(\theta, x, y) \cos 2\phi,
\]
where
\[
\begin{split}
Q(\theta, x, y) &= (1-x-y)\sin^2\theta + y_d(1+\sin^2\theta) + x_d\cos^2\theta \\
&\quad - \frac{1}{8}\sin 2\theta \left[(1-x-y)^2 + (y_d - x_d)^2\right],
\end{split}
\]
and
\[
R(\theta, x, y) = 2\sin^2\theta \cos^2\theta (1-x-y) (y_d - x_d).
\]
Since  $\lambda_1+\lambda_2$ is independent of the input state, the entropy is minimized if we minimize $\lambda_1\lambda_2$ or the determinant of $M$. To do this, we note the following two separate conditions. 

\subsubsection{Holevo information in the region $y_d\geq x_d$} This is the region $A$ shown in figure (\ref{fig:Phases}).
For this region of the parameters $x$ and $y$, $R(\theta,x,y)\geq 0$ and hence the $Det(M)$ is minimized if we set $\phi=0$. 
Direct calculation then shows that the $Det(M)$ is indeed independent of $\theta$ and is equal to
\be
\det M = \lambda_1\lambda_2=(x_d+y_d)\Big[1-(d-1)(x_d+y_d)\Big]
\ee
Given the sum of eigenvalues to be the one given in (\ref{sumeig}), we find

\be
\lambda_1=1-(d-1)(x_d+y_d)\h \lambda_2=y_d+x_d\h \lambda_i=\lambda_2 \ , \forall\ \ i=3\cdots d.
\ee
Therefore the Holevo information in this region becomes 
\be\label{capA}
C^1(\Phi_{_{x,y}})=\log d+\lambda_1\log \lambda_1+(d-1)\lambda_2\log \lambda_2
\ee
In this case any state of the form 
\be
\phi=0\lo |\psi\ra=\begin{pmatrix}\cos\theta \\ \sin\theta \\ 0 \\ \cdot \\ 0\end{pmatrix},
\ee
is a minimum output entropy state regardless of the value of $\theta$. This means that all the states connected by $SO(2)\times SO(d-2)$ rotations are minimum output entropy states. 
\subsubsection{Holevo information in the region $y_d< x_d$} 
\begin{figure}[H]
	\centering
	
	\includegraphics[width=\textwidth]{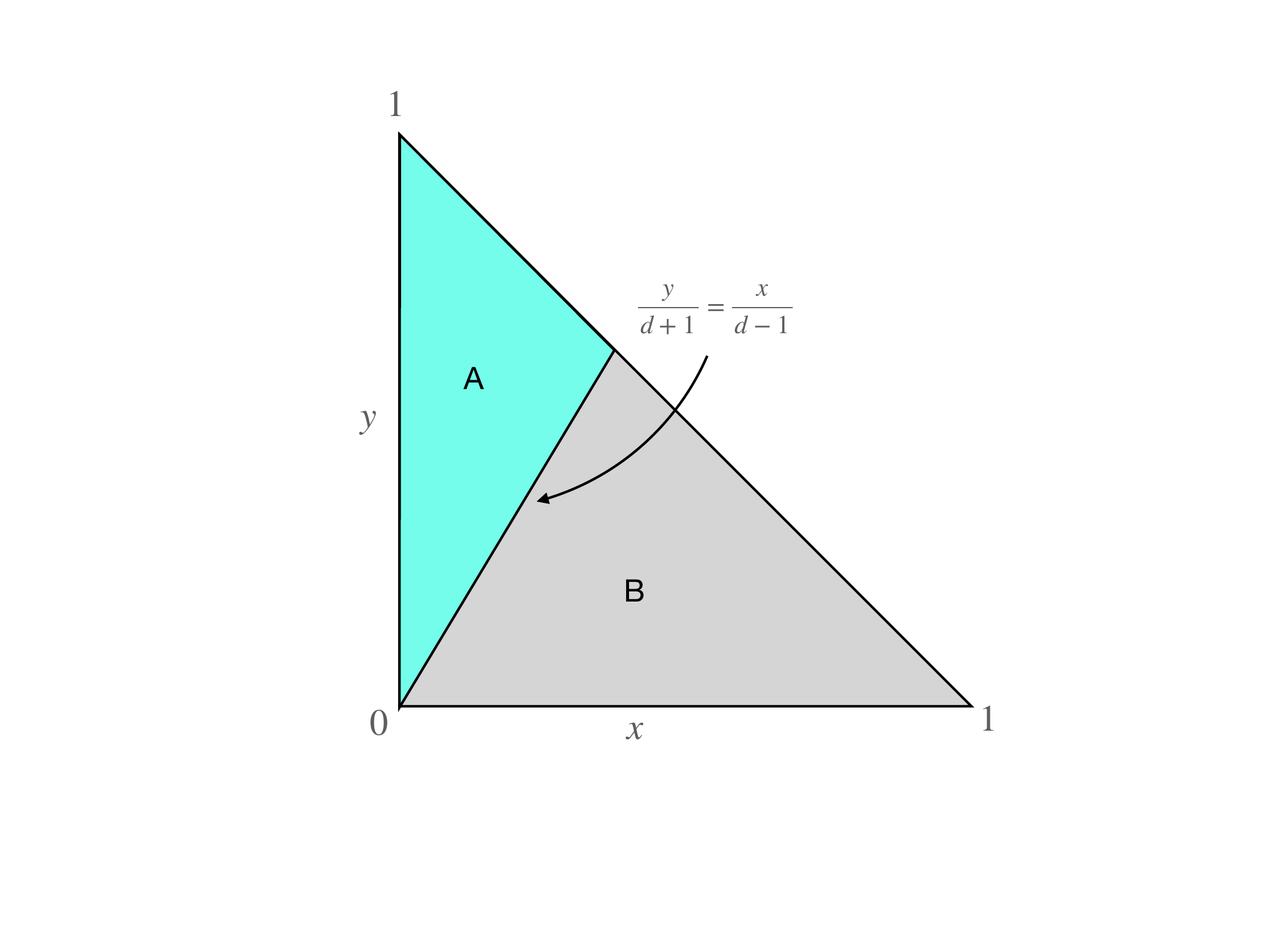}\vspace{-2.5cm}
	
	\caption{The classical one-shot capacity in the two regions $A$ and $B$, given respectively in equations (\ref{capA}) and (\ref{capB}). }
	\label{fig:Phases}
\end{figure}

This is the region $B$ shown in figure (\ref{fig:Phases}).
In this region, $R(\theta,x,y)$ is negative and we can minimize $Det(M)$ by setting $\phi=\frac{\pi}{2}$.  Calculation then sees that the minimum of $Det(M)$ is achieved at $\theta=\frac{\pi}{4}$. The minimum output entropy state is now of the form
\be
\theta=\frac{\pi}{4}\lo |\psi\ra=\frac{1}{\sqrt{2}}\begin{pmatrix}i \\ 1 \\ 0 \\ \cdot \\ 0\end{pmatrix}.
\ee
$Det(M)$ turns out to be	
\be
\det M = \mu_1\mu_2=2y_d\Big[1-(d-2)x_d-dy_d\Big]
\ee
In conjunction with (\ref{sumeig}), this gives the eigenvalues to be
\be
\lambda'_1=2y_d\h \lambda'_2=1-(d-2)x_d-dy_d,\h \lambda'_i=y_d+x_d\ , \forall \ i=3,\cdots d
\ee
Leading to the following value for the classical one-shot capacity		
\be\label{capB}
C^1(\Phi_{_{x,y}})=\log d+\lambda'_1\log \lambda'_1+\lambda'_2\log \lambda'_2+(d-2)\lambda'_3\log \lambda'_3
\ee
Using equations (\ref{capA}) and (\ref{capB}), one can check that the function $C^{(1)}$ is continuous inside the whole triangle $x+y\leq 1. $
The one-shot classical capacity is not differentiable in the parameter space, but it may smooth out when one considers the regularized classical capacity. The analytical calculation of this capacity is however is not possible.

\subsection{ Bounds for the Classical Capacity}
\label{CC-UB}
One can find an upper bound for classical capacity of channel $\Lambda: A\rightarrow B$ using the following relation \cite{wang_classical}:
$$C(\Lambda)\le \log(d_B   \Vert  J(\Lambda)^{T_B}\Vert _\infty),$$
where $J(\Lambda)=\sum_{ij}\dyad{i}{j}\otimes \Lambda(\dyad{i}{j})$ is the Choi matrix of the channel \cite{choi1975}. $A^{T_B}$ corresponds to partial transpose with respect to subspace $B$. $\abs{\abs{A}}_\infty$, corresponds to maximum singular value of matrix $A$.
 Note that $J(\Phi_{x,y})^{T_B}$ is a hermitian matrix, as a result its singular values are nothing but absolute value of its eigenvalues. Now we first determine the eigenvalues of $J(\Phi_{x,y})^{T_B}$. With straightforward calculations one can see that:
\begin{equation}
	J(\Phi_{x,y})=(1-x-y)\sum_{ij} \dyad{ii}{jj}+(y_d+x_d)I+(y_d-x_d)\sum_{ij}\dyad{ij}{ji}
\end{equation}
Hence,
\be	
J(\Phi_{x,y})^{T_B}=(1-x-y)P+(y_d+x_d)I+d(y_d-x_d)|\phi^+\ra\la \phi^+|,
\ee
where $\ket{\phi^+}=\frac{1}{\sqrt{d}}\sum_i{\ket{ii}}$ and $P$ is the permutation operator $P|i,j\ra=|j,i\ra$. The three operators above commute with each other which facilitates the determination of the spectrum. One immediate eigenvectors is $|\phi^+\ra$ with eigenvalue equal to the sum of the above three coefficients which turns out to be equal to $1-2x$. The other eigenvectors are those which are orthogonal to $|\phi^+\ra$ and have specific symmetries under interchange of the two vectors. The complete spectrum (eigenvalues and eigenvectors) are:

\be\label{spectb} 	\text{Spectrum}(J(\Phi_{x,y})^{T_B})=
\begin{cases}
	1-2 x & \ \ \ \ |\phi^+\ra \\
	1-x-y+y_d+x_d,  & \ \ \ \ |i,j\ra+|j,i\ra\ \ \ \h \ \ i<j \\
	-(1-x-y)+y_d+x_d,  &\ \ \ \   |i,j\ra-|j,i\ra\ \ \ \h \ \  i<j \\
	(1-x-y)+y_d+x_d,  &\ \ \ \   |i,i\ra-|i+1,i+1\ra\ \ \ \ i=1 \cdots d-1.
\end{cases}
\ee
As $1\ge x+y$ we easily see that :
\be
\Vert{J(\Phi_{x,y})^{T_B}}\Vert_\infty=\max\{1-x-y+y_d+x_d,  \abs{1-2x}\}.
\ee
This gives the following expression for an upper bound of classical capacity of the channel $\Phi_{x,y}$:\\

\noindent {\bf For $x\leq \frac{1}{2}$:}

\be
C_{cl}(\Phi_{x,y})\leq 	\begin{cases}
\log d+\log \big[	1-x-y+y_d+x_d \big]& \text{if}\quad  x_d  > y_d \\
\log d +\log 	\big[1-2x \big] & \text{if} \quad  x_d  \le y_d 
\end{cases}
\ee
 
\noindent  {\bf For $x> \frac{1}{2}$:}

 \be
 C_{cl}(\Phi_{x,y})\leq \begin{cases}
 \log d+\log \big[	1-x-y+y_d+x_d\big] & \text{if}\quad 2 > dy_d+3x-x_d \\
 	\log  d+\log \big[2x-1 \big] & \text{if}\quad 2   \leq  dy_d+3x-x_d
 \end{cases}
 \ee
 where all the logarithms are in base $2$. \\
 
\begin{figure}[H]
\centering
\begin{subfigure}{0.48\textwidth}
  \centering
  \includegraphics[width=\linewidth]{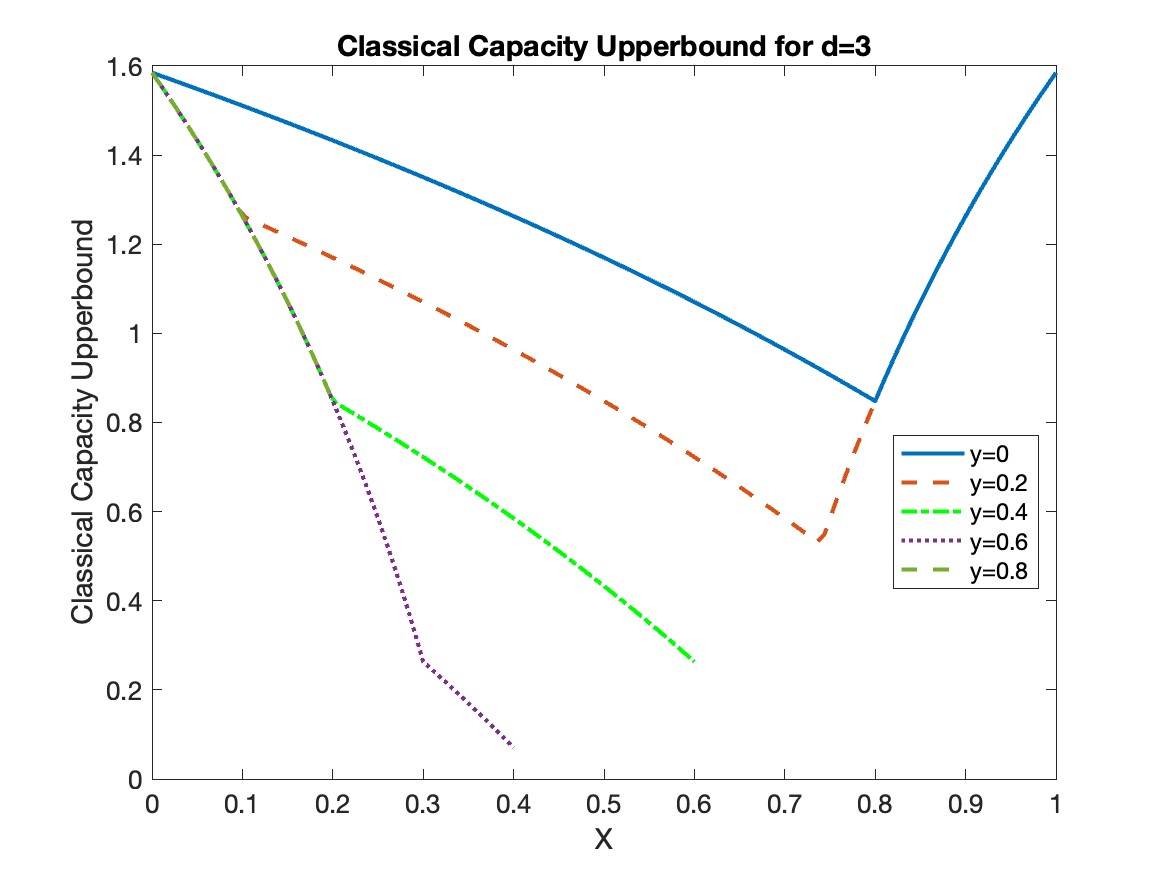}
  %\caption{Caption for image 1}
  \label{fig:sub1}
\end{subfigure}
\begin{subfigure}{0.48\textwidth}
  \centering
  \includegraphics[width=\linewidth]{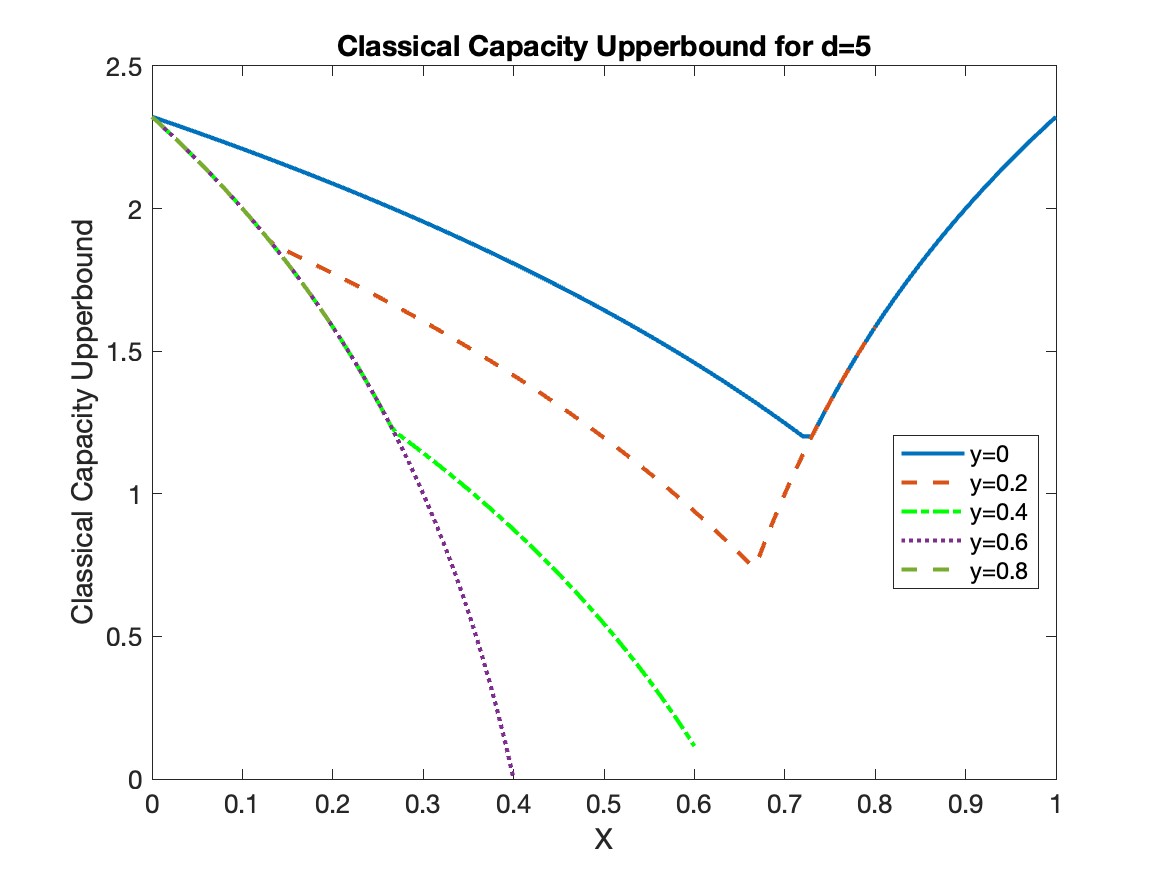}
  %\caption{Caption for image 2}
  \label{fig:sub2}
\end{subfigure}

\begin{subfigure}{0.48\textwidth}
  \centering
  \includegraphics[width=\linewidth]{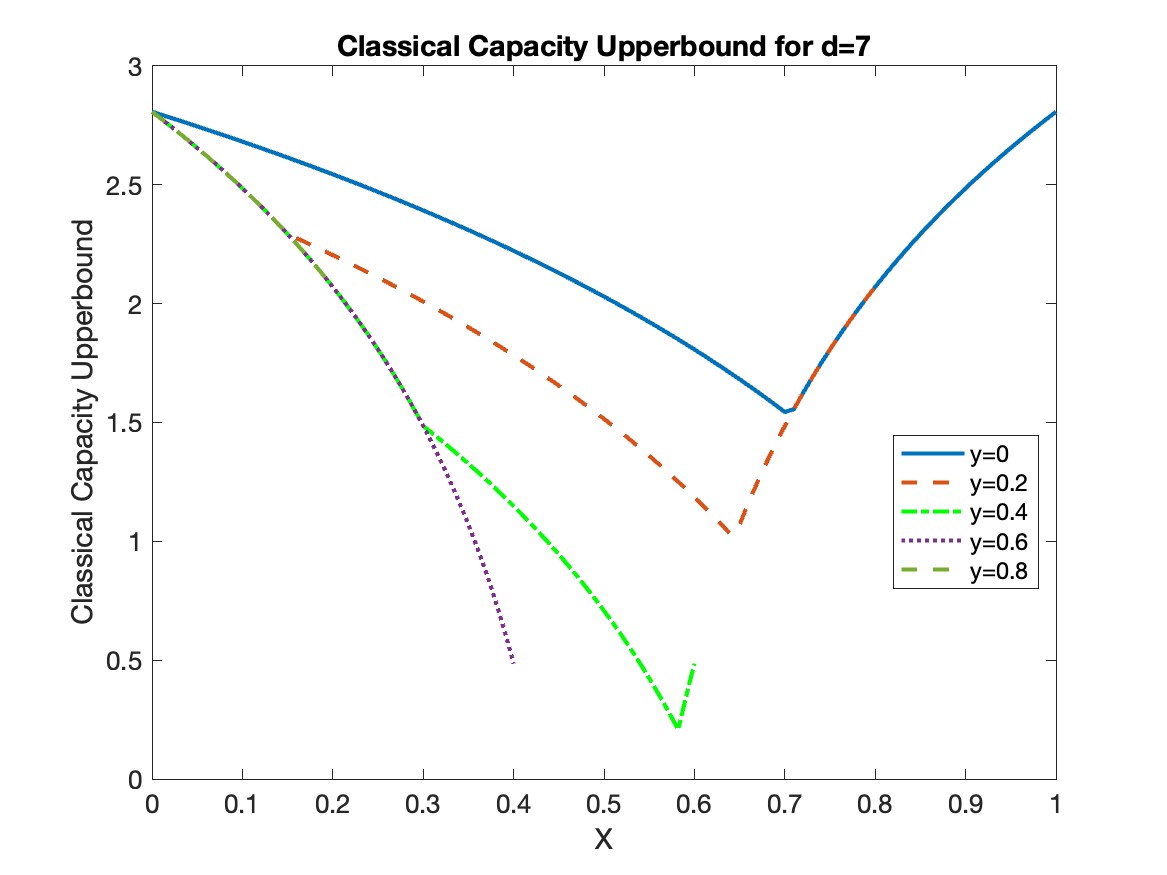}
  %\caption{Caption for image 3}
  \label{fig:sub3}
\end{subfigure}
\begin{subfigure}{0.48\textwidth}
  \centering
  \includegraphics[width=\linewidth]{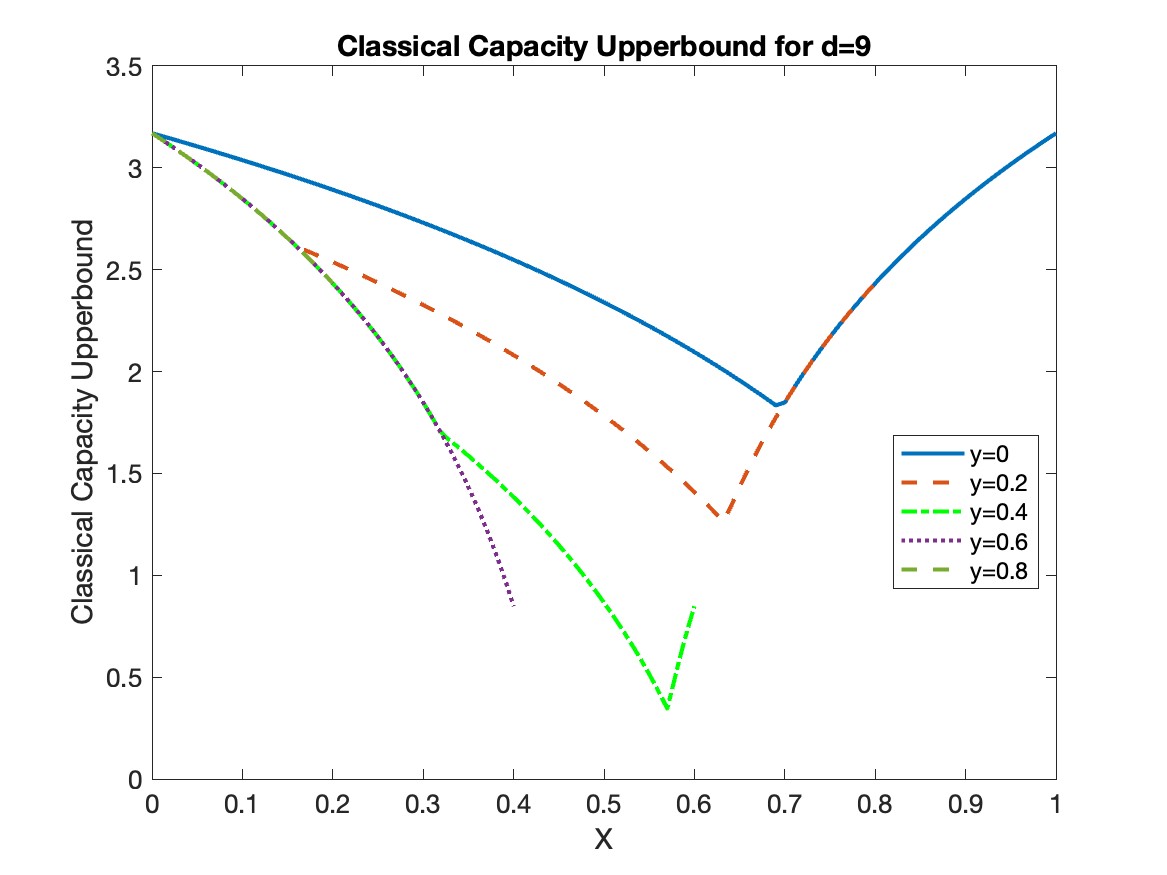}
 % \caption{Caption for image 4}
  \label{fig:sub4}
\end{subfigure}
\caption{The figure displays upper bounds of classical capacity for channels with dimensions $d=3,5,7,9$ and for varying values of $y$ (0, 0.2, 0.4, 0.6, and 0.8), depicted as a function of $x$. }
\label{fig:CC-UB}
\end{figure}

\ni  The results on classical capacity are shown in figures (\ref{fig:CC-UB}) for several values of dimension $d$. 

\section{Entanglement-Assisted Classical Capacity}\label{ent}
Quantum dense coding is a paradigmatic example of a quantum communication protocol in which an unlimited number of maximally entangled pairs are at the disposal of the sender and reciever. When the quantum channel between Alice and Bob is noiseless, they can communicate two bits of classical information by sending just one qubit (i.e. one qubit from the entangled pair after suitable measurement.) Therefore the entanglement-assisted classical capacity of a noiseless channel is $2$ bits per use of the channel. the question is what this classical capacity is, if the quantum channel is noisy. A closed formula for this quantity has been provided in  \cite{bennett_entanglement-assisted_1999} and for a general quantum channel $\Lambda$ is given by \cite{bennett_entanglement-assisted_2002}:
\begin{equation}
	C_{ea}(\Lambda)=\max_{\rho} I(\rho,\Phi)
\end{equation}
where 
\be
I(\rho, \Lambda):=S(\rho)+S( \Lambda(\rho))-S(\rho, \Lambda).
\ee
Here $S(\rho,\Lambda)$ is the output entropy of the environment, referred to as the entropy exchange \cite{nielsen_1998}, and is represented by the expression $S(\rho,\Lambda) = S\left(({\Lambda}\otimes I)  \Psi_\rho\right)$ where $\Psi_\rho$ is a purification of the state $\rho$. A purification of $\rho$ can be obtained by $\Psi_\rho= \dyad{\sqrt{\rho}}$, where $\ket{A}=\sum_{ij} A_{ij} \ket{ij}$ is the vectorized form of the matrix  $A=\sum_{i,j}A_{i,j}|i\ra\la j|$.\\

\ni According to proposition 9.3 in \cite{holevoBook}, the maximum entanglement-assisted capacity of a covariant channel  is attained for an invariant state $\rho$. In the special case where the channel is irreducibly covariant, the maximum is attained on the maximally mixed state.
Hence, for the channel $\Phi_{_{x,y}}$, we have 
\begin{equation}
	C_{ea}(\Phi_{_{x,y}})=S(\frac{I}{d})+S(\Phi_{_{x,y}}(\frac{I}{d}))-S((\Phi_{_{x,y}}\otimes I)  \Psi_{\frac{I}{d}}),
\end{equation}
which, given the unitality of the channel and the fact that a purification of the maximally mixed state is maximally entangled state , lead us to
\be
	C_{ea}(\Phi_{_{x,y}})=2S(\frac{I}{d})-S((\Phi_{_{x,y}}\otimes I) \: \dyad{\phi^+})=2\log_2d-S(\frac{1}{d}J_{\Phi_{x,y}}),
	\label{eq1}
\ee
where $J_{\Phi_{x,y}}$ is the Choi matrix of the channel. From (), this Choi matrix is seen to be equal to
\be
J(\Phi_{x,y})=(1-x-y)d|\phi^+\ra\la \phi^+|+(x_d+y_d)I+(y_d-x_d)P
\ee
where $P$ is the permutation operator. The eigenvalues of this matrix are easily evaluated (by the same type of reasoning as was used in section (\ref{CC-UB}) for the matrix $J(\Phi_{x,y})^{T_B}$). The result is

\be 	\text{Spectrum}(\frac{1}{d}J(\Phi_{x,y}))=
\begin{cases}
	(1-x-y)+\frac{2y_d}{d} & \ \ \ \ |\phi^+\ra \\
	\frac{2y_d}{d},  & \ \ \ \ |i,j\ra+|j,i\ra\ \ \ \h \ \ i<j \\
	\frac{2x_d}{d},  &\ \ \ \   |i,j\ra-|j,i\ra\ \ \ \h \ \  i<j \\
	\frac{2y_d}{d},  &\ \ \ \   |i,i\ra-|i+1,i+1\ra\ \ \ \ i=1 \cdots d-1.
\end{cases}
\ee
Therefore we find
\be
	C_{ea}(\Phi_{_{x,y}})=2\log d+ \xi \log \xi + \frac{(d-1)(d+2)}{d} y_d\log \frac{2y_d}{d}+(d-1) x_d\log \frac{2x_d}{d},
	\label{eq2}
\ee
where $\xi=(1-x-y)+\frac{2y_d}{d}$. 

\section{Quantum Capacity}
\label{QuantumCapacity}
This is the ultimate rate for transmitting quantum information and preservation the entanglement between the channel's input and a reference quantum state over a quantum channel. This quantity is described in terms of coherent information \cite{devetak_private_2005}:

\begin{equation}
	\begin{split}
		Q(\Lambda)= \lim_{n \to \infty} Q_n (\Lambda) =
		\lim_{n \to \infty} \; \max_{\rho} \; \frac{1}{n} J(\rho ,
		\Lambda^{\otimes n}),
	\end{split}     
\end{equation}
where $J(\rho,\Lambda)=S(\Lambda(\rho))-S(\Lambda^c(\rho))$ is superadditive (i.e. $Q_1(\Lambda)\leq Q(\lambda)$) and non-convex in general. However, it turns out that  for  degradable channels, $Q(\Lambda)=Q_1(\Lambda)$ and the calculation of the Quantum capacity becomes a convex optimization problem. On the other hand if the channel is anti-degradable, it is known that its quantum capacity vanishes \cite{smith_detecting_2012}. In \cite{roofeh_noisy_2023} we showed that in $d=3$ dimension, the channel $\Phi_{x,0}$ is antidegradable when $\frac{4}{7}\leq x\leq 1. $ It is also known that if a channel is entanglement breaking, then its quantum capacity is zero. These two regions are not necessarily the same, as was shown for example in the case of Pauli channel in \cite{poshtvan_capacities_2022}. An entanglement-breaking channel has the property that its Choi matrix is separable. However in higher dimensions, checking separability of the Choi matrix is itself a difficult problem. There are however other techniques which provide upper bounds for the quantum capacity \cite{kianvash1, kianvash2,fern_lower_2008,wang_2016}. In this section, we use these techniques to find upper bounds for the quantum capacity. It turns out that in certain regions of the parameter space, these this upper bound is zero, which implies that these regions are the zero-capacity region. A natural lower bound for the quantum capacity is the single-shot capacity defined as $C_q^{(1)}(\Phi_{x,y})=max_{\rho} J(\rho, \Phi_{x,y})$, where $J$ is the coherent information. 

\subsection{Zero-region Capacity}
\label{Zero-region Capacity}
While it is generally challenging to find tight upper bounds for quantum capacity, it is possible by moderate effort to determine regions of parameters where the quantum capacity is exactly zero. When dealing with qubit channels, there are two well-known techniques for doing this. First, one can determine the region where the channel is entanglement-breaking for which it is enough to determine the Choi matrix and apply the Peres criteria \cite{peres_separability_1996} to see if it is seperable or not. The separability of the Choi matrix is equivalent to entanglement-breaking property of the channel. Such a channel completely destroys any initial entanglement and hence its quantum capacity vanishes. Another method is to see under what conditions the channel is anti-degradable, where again for qubits, there is a simple criteria based on the Choi matrix. Namely, according to \cite{paddock_characterization_2017}, a qubit channel $\Lambda$ is anti-degrdable if its Choi-matrix ${\cal J}_{\Lambda}$ satisfies the following inequality
\be
\tr( {\cal J}_{\Lambda})^2-4\sqrt{det( {\cal J}_{\Lambda})}\leq \tr(\Lambda(I)^2).
\ee 
The regions determind by these two methods are not necessarily identical and both of them together determine a region of zero-quantum-capacity. In \cite{poshtvan_capacities_2022}, both methods have been used to determine this region for a covariant two-parameter qubit Pauli channel.  We should stress that the actual zero-quantum-capacity may be larger than the one obtained in this way. \\
When we go to higher dimensions, there is no closed form criteria for anti-degradability of the channel and the set of separable states is not equal to PPT states. However, it is known that entanglement-binding channels \cite{horodecki_separability_1997,horodecki_separability_1996}, also called PPT channels have zero quantum capacity. These channels have Choi matrices with positive partial trace and can only produce very weakly entangled states satisfying PPT conditions \cite{peres_separability_1996,smith_quantum_2008,horodecki_separability_1997,horodecki_separability_1996}. 

\begin{figure}[H]
	\centering
	\includegraphics[width=12cm]{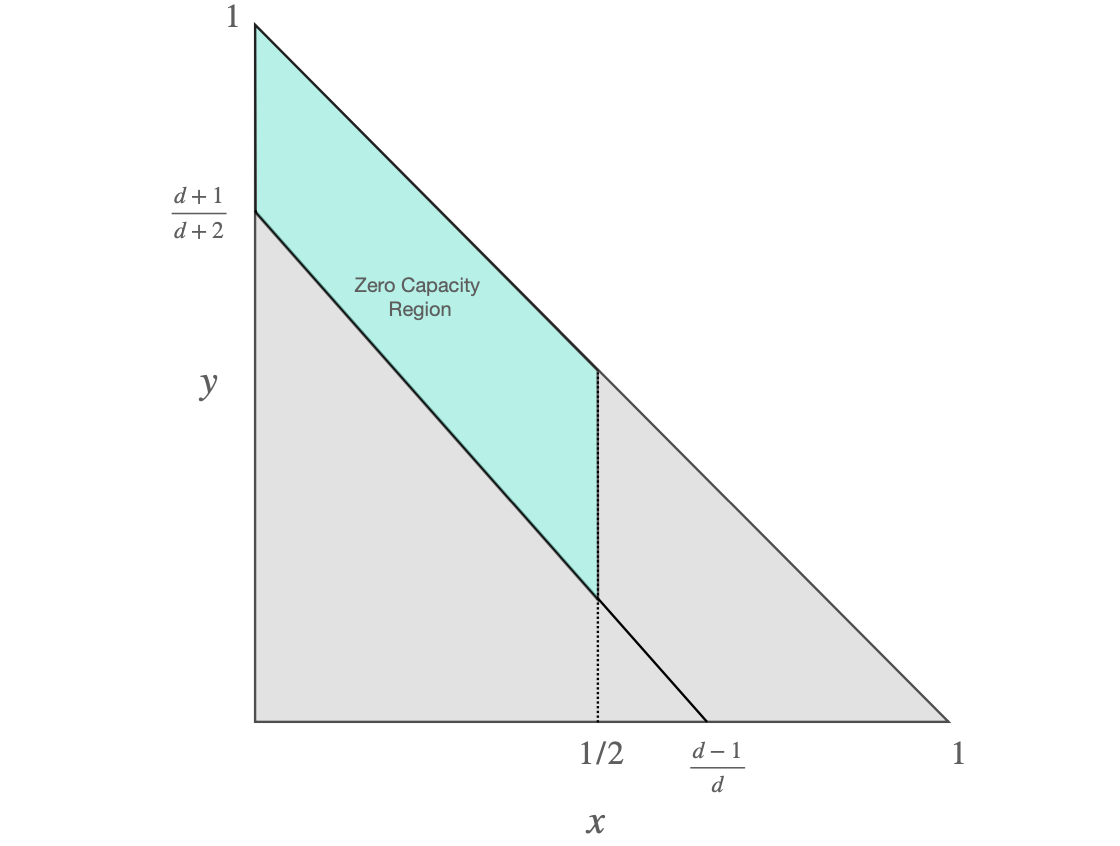}\vspace{0 cm}
	\caption{The region where the quantum capacity of the channel is zero. See the text for a proof. }
	\label{ZCR}
\end{figure}

\ni {\bf Theorem:} The quantum capacity of the channel $\Lambda_{x,y}$ is zero in the region $$\Omega_d =\{0\leq x\leq \frac{1}{2}, \ \ \ 1\leq d x_d+(d+2)y_d\}$$.\\
\ni {\bf Proof:}
The spectrum of $J_\Lambda^{T_B}$ has already  been obtained in (\ref{spectb}). It is readily seen that
 $J_\Lambda^{T_B}\geq 0$ in the region $\Omega_d$. Hence, the channel is PPT and has zero-capacity in this region.\\
This region is shown  in figure (\ref{ZCR}).
Note that in some region of the parameters, the channel may be entanglement-breaking and its quantum capacity can be zero. This region is recognized by demanding that the Choi matrix ${\cal J}_\Lambda$ be separable. Therefore in this region, the partial transpose of the Choi matrix is positive, that is ${\cal J}_\Lambda^{T_B}\geq 0$. This means that the entanglement breaking region is a subset of the region $\Omega_d$.\\

\ni  {\bf Example: the case of} $d=2$\\

\ni  When $d=2$, we have 
$$\Omega_2=\{0\leq x\leq \frac{1}{2},\ \ \  \frac{1}{2}\leq x+\frac{2}{3}y\}.$$ In this dimension, the channel $\Lambda_{x,y}$ acts as 
\be
\Lambda_{x,y}(\rho)=1-x-y+ x(\tr(\rho) I-\rho^T)+\frac{y}{3}(\tr(\rho)I+\rho^T).
\ee
Since in two dimensions, we have $\tr(\rho)I-\rho^T=\sigma_y\rho \sigma_y$ and $\sum_{\mu=0}^3\sigma_\mu\rho \sigma_\mu=2\tr(\rho)I$ with $\sigma_\mu=(I,\sigma_x,\sigma_y,\sigma_z)$, one finds after a simple algebra that
\be
\Lambda_{x,y}(\rho)=(1-x-\frac{2}{3}y)\rho+ \frac{y}{3}\big(\sigma_x\rho \sigma_x +\sigma_z\rho \sigma_z\big)+x\sigma_z\rho \sigma_z.
\ee
This is the two parameteric covariant Pauli channel 
\be
\Phi(\rho)=p_0\rho + p_1\big(\sigma_x\rho \sigma_x +\sigma_z\rho \sigma_z\big)+p_3\sigma_y\rho \sigma_y.
\ee
which was recently analyzed in \cite{poshtvan_capacities_2022}, where using the entanglement-breaking condition of \cite{peres_separability_1996}, it was   shown that the zero-capacity region covers the area
\be
\{p_0\leq \frac{1}{2},\ \ \ \ p_3\leq \frac{1}{2}\}.
\ee
Expressing $p_0$ and $p_3$ in terms of $x$ and $y$, i.e. $(p_0=(1-x-\frac{2}{3}y),\ \ p_3=x)$, we see that this region coincides with $\Omega_2$. We emphasize that the actual zero-capacity region may be larger than $\Omega_d$. 

\subsection{ Upper bound }
One can use techniques induced from semi-definite programming to determine parts of this region. In fact, an upper bound for quantum capacity of a channel by semi-definite programming is introduced in ref.\cite{wang_2016} and is denoted by $Q_\Gamma$. Let $\Lambda: A \rightarrow B$ be a quantum channel, and 
${\cal J}_{\Lambda}= \sum_{ij} (\dyad{i}{j}) \otimes (\Lambda(\dyad{i}{j}))$ be its Choi-matrix. Also let 
$\rho_A$ be an arbitrary density matrix in $A$, and $R$ be an arbitrary positive semi-definite linear operator in $L^+(A\otimes B)$. 
Then the upper bound for the quantum capacity is found as follows \cite{wang_2016}
\be
\label{SDP-Upperbound}
    Q(\Phi) \leq Q_\Gamma(\Lambda) := \log\left(\max_R \:\operatorname{Tr}({\cal J}_{\Lambda}R)\right),
\ee    
where $R$ is a positive operator in $L(A\otimes B)$ subject to the following conditions:
\be
  -\rho\otimes I \leq R^{T_B} \leq \rho\otimes I,\h R, \rho\geq 0,\ \ \ \tr(\rho)=1.
\ee
Here \(T_B\) represents the partial transpose operation with respect to space \(B\), defined as \((\dyad{ij}{kl})^{T_B} = \dyad{il}{kj}\).\\

\begin{figure}
\centering
\begin{subfigure}{0.45\textwidth}
  \centering
  \includegraphics[width=\linewidth]{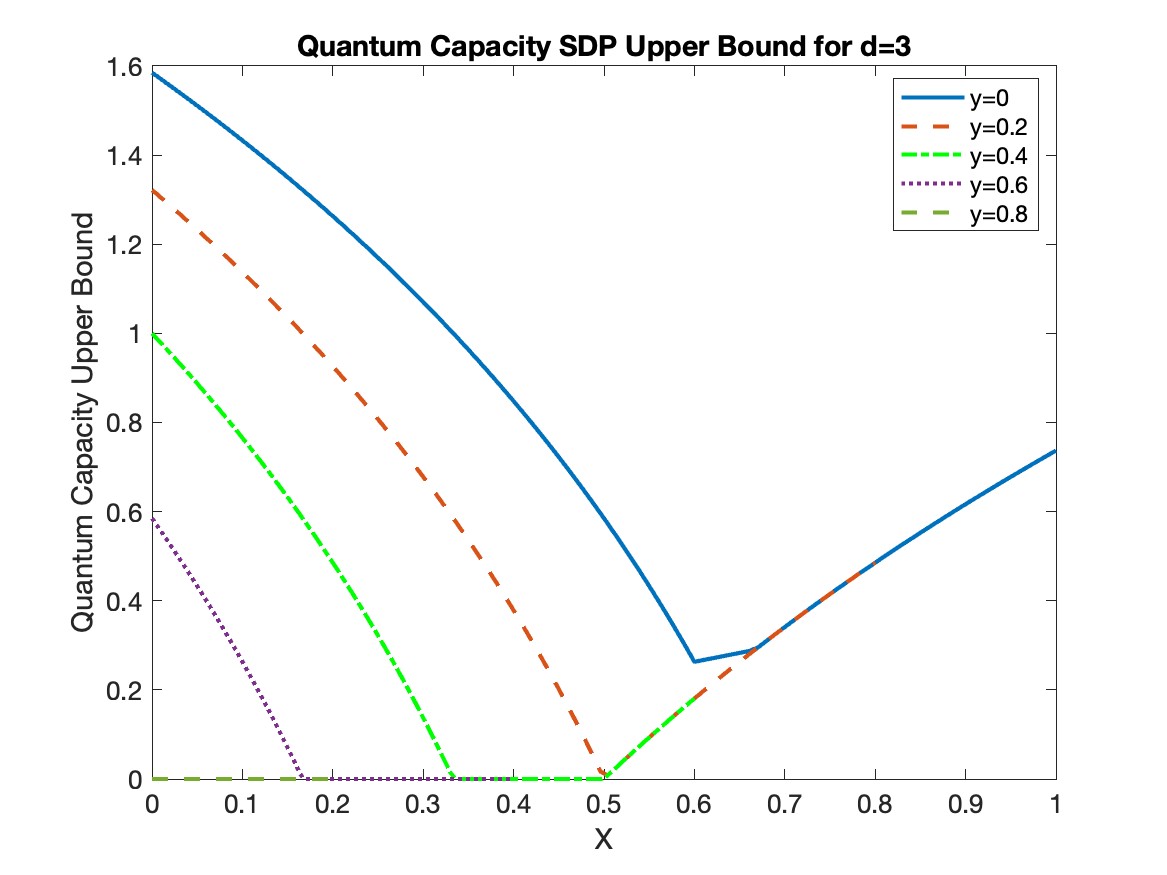}
  %\caption{Caption for image 1}
  \label{fig:sub1}
\end{subfigure}
\begin{subfigure}{0.45\textwidth}
  \centering
  \includegraphics[width=\linewidth]{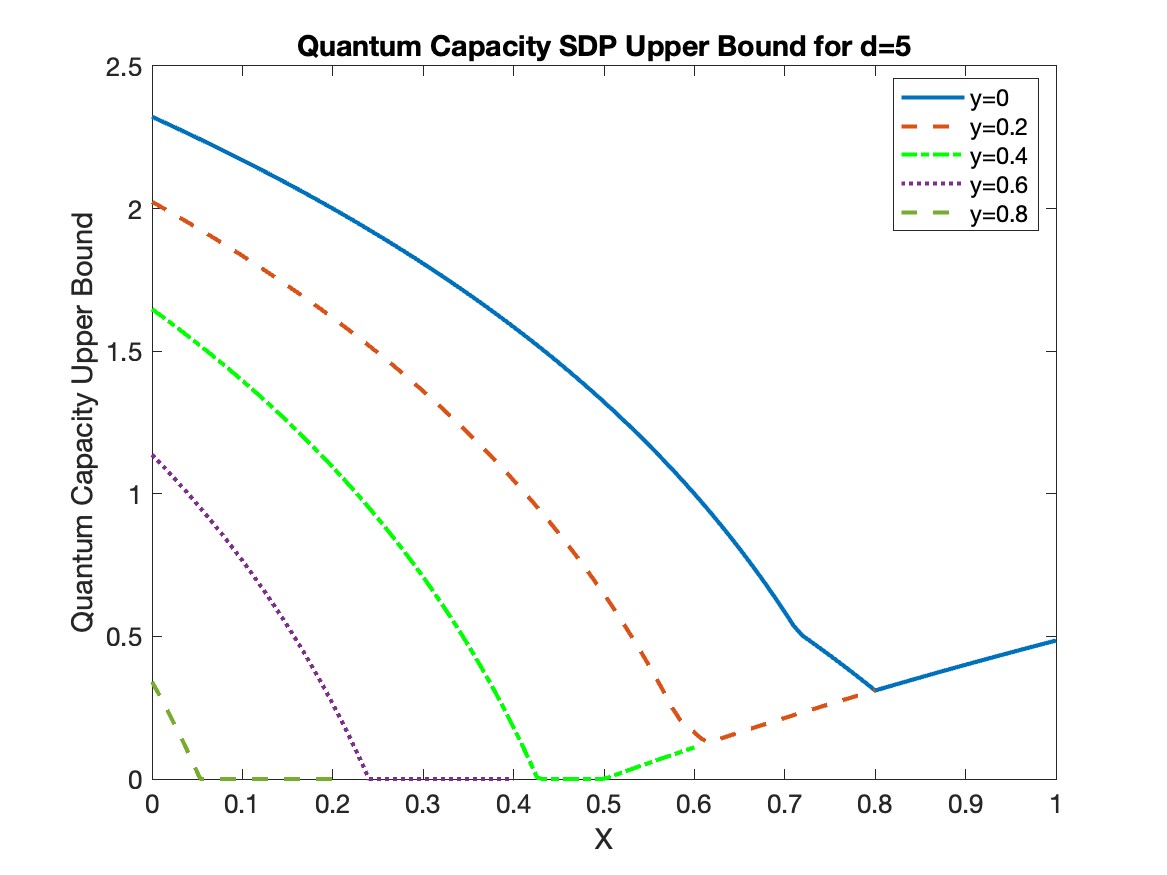}
  %\caption{Caption for image 2}
  \label{fig:sub2}
\end{subfigure}

\begin{subfigure}{0.45\textwidth}
  \centering
  \includegraphics[width=\linewidth]{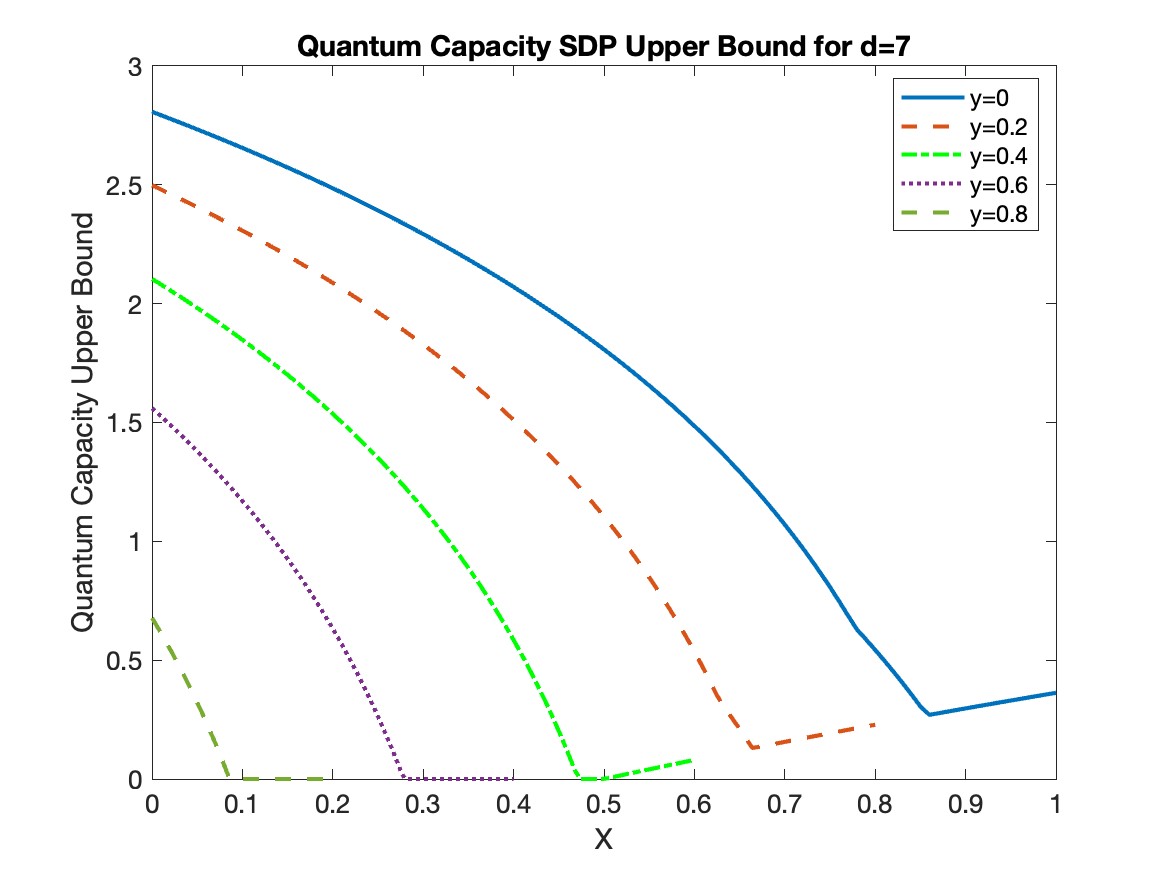}
  %\caption{Caption for image 3}
  \label{fig:sub3}
\end{subfigure}
\begin{subfigure}{0.45\textwidth}
  \centering
  \includegraphics[width=\linewidth]{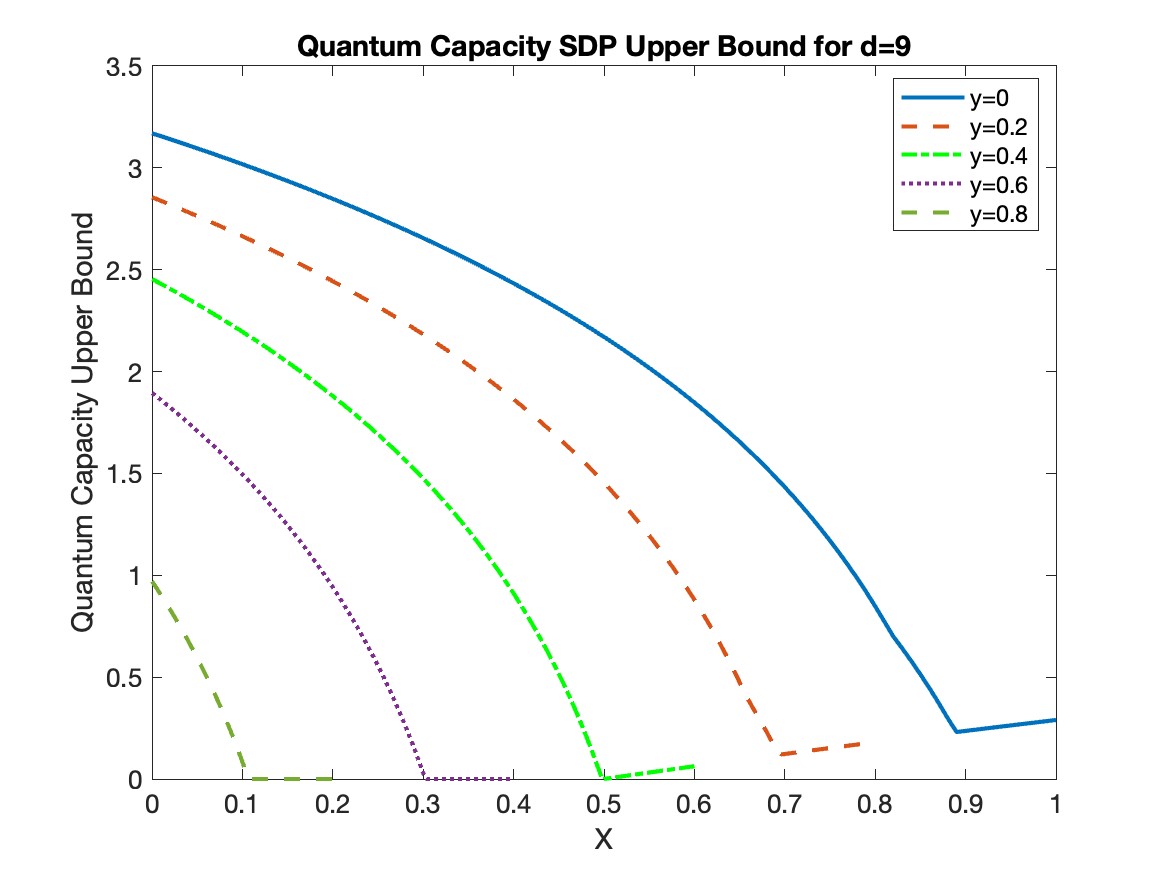}
 % \caption{Caption for image 4}
  \label{fig:sub4}
\end{subfigure}
\caption{The figure displays upper bounds for channels with dimensions $d=3,5,7,9$ for varying values of $y$ (0, 0.2, 0.4, 0.6, and 0.8), depicted as a function of $x$. Agreement with the zero capacity region in figure (\ref{ZCR}) is evident. }
\label{SDP-UB}
\end{figure}

By using the semi-definite program (\ref{SDP-Upperbound}) , we can now numerically establish upper bounds for each choice of $x$, $y$ and $d$. Figure \ref{SDP-UB} illustrates the different upper bounds for $d=3,5,7,9$ with various values of $y$, presented as a function of $x$. As it is evident, the results agree with the zero quantum capacity region evaluated in section \ref{Zero-region Capacity}. In fact the solution of SDP program (\ref{SDP-Upperbound}) for PPT channels is zero. To show this, first observe that $\tr(RJ_\Lambda)=\tr(R^{T_B}J^{T_B}_\Lambda)$. Also,
 $J_\Lambda^{T_B}\geq 0$ in the region $\Omega_d$ where the channel is PPT. Like any other channel, it is also known that  $\Tr( J_\Lambda^{T_B}(\rho\otimes I))=\Tr\Lambda(\rho)=1$. 
 In view of positivity of $J_\Lambda^{T_B}$ in $\Omega_d$, this means that for any matrix $R$ satisfying $R^{T_B}\leq \rho\otimes I$,
the maximum of $\Tr(J^{T_B} R^{T_B})$ is at most equal to unity which implies that $Q(\Lambda)\equiv \log\left(\max_R \:\operatorname{Tr}({\cal J}_{\Lambda}R)\right)=0$. Therefore in the region $\Omega_d $, the quantum capacity of $\Lambda_{x,y}$ is equal to zero.

\subsection{Lower bound}
A natural lower bound for the quantum capacity is given by single-shot quantum capacity $Q^1(\Phi_{_{x,y}})$, defined as 
 \cite{barnum_information_1998, lloyd_capacity_1997,devetak_private_2005}:
\begin{equation}
	\begin{split}
		Q^{(1)}(\Lambda) =
		max_{\rho}J(\rho,
		\Lambda),
	\end{split}     
\end{equation}
where $J(\rho,\Lambda):=S(\Lambda(\rho))-S(\Lambda^c(\rho))$ is the coherent information. This lower bound owes its existence to the superadditivity of the coherence information, i.e. to the inequality  $J(\Lambda_1 \otimes \Lambda_2)\ge J(\Lambda_1)+J(\Lambda_2)$.  For some quantum channels which are degradable, the additivity property is restored $Q(\Lambda)=Q^{(1)}(\Lambda)$ \cite{devetak_capacity_2005}, and the calculation of the quantum capacity becomes a convex optimization problem. In view of the definition single shot capacity of the channel $\Phi_{_{x,y}}$ as given by  
\be
Q^{(1)}(\Phi_{_{x,y}})=Max_{\rho}J(\rho,\Phi_{_{x,y}})=Max_{\rho}\Big[S(\Phi_{_{x,y}}(\rho))-S(\Phi_{_{x,y}}^c(\rho))\Big],
\ee
we know that  any state $\rho$, not necessarily the one which maximizes the coherent information will also provide a lower bound for the quantum capacity although it may not a tight bound. 
Therefore we choose $\rho$ to be simply the maximally mixed state $\rho=\frac{I}{d}$. This will give 
\be
\label{eq3}
Q^{(1)}(\Phi_{_{x,y}})\ge \Big[S(\Phi_{_{x,y}}(\frac{I}{d}))-S(\Phi_{_{x,y}}^c(\frac{I}{d}))\Big]
\ee
which in view of (\ref{eq1}) turns out to be simply related to the entanglement-assisted capacity. Therefore we use equation (\ref{eq3}) and find a lower bound in the form $ Q^{(1)}(\Phi_{_{x,y}})=C_{ea}(\Phi_{_{x,y}})-\log d$ or 
 \be
 Q^{(1)}(\Phi_{_{x,y}})=\log d+ \xi \log \xi + \frac{(d-1)(d+2)}{d} y_d\log \frac{2y_d}{d}+(d-1) x_d\log \frac{2x_d}{d}
 \ee
 where as in (\ref{eq2}), $\xi=(1-x-y)-\frac{2y_d}{d}.$

\section{Discussion}
We have generalized the Landau-Streater channel which is pertaining to the spin-j representation of the Lie-algebra $su(2)$ to the fundamental representation of the groups $u(d)$ and $su(d)$ and have pointed out their equivalence to the Werner-Holevo channels. We have studied several properties of the resulting two-parameter family of quantum channels, including their complement channels,  their one-shot classical capacity and their entanglement-assisted classical capacity. In particular, we have determined in closed form an area in the space of parameters, where the quantum capacity vanishes. Besides that, by using semi-definite programming, we have obtained lower and upper bounds for the quantum capacity of this channel in its full range of parameters.

	\section{Acknowledgements} 
We would like to thank members of the QIS group in Sharif, especially V.Jannesari, A. Farmamian, and A. M. Tangestani for their valuable comments.

\newpage

\bibliography{refs}

\newpage

\end{document}